\documentclass[10pt, conference, letterpaper]{IEEEtran}
\IEEEoverridecommandlockouts
  
\usepackage[utf8]{inputenc}
\usepackage{makeidx}
\makeindex
\usepackage[T1]{fontenc}
\usepackage[pdftex]{graphicx}
\usepackage{epstopdf}
\usepackage{cite}
\usepackage{url}
\usepackage[font=footnotesize]{caption}
\usepackage{color}
\usepackage{amsmath}
\usepackage{amssymb}
\usepackage{multirow}
\usepackage{eso-pic}
\usepackage{balance}
\usepackage{listings}
\usepackage{xspace}
\usepackage{subcaption}

\usepackage{acro}

\DeclareAcronym{gui}{short = GUI, long = graphical user interface}
\DeclareAcronym{aa}{short = AA, long = authentication and authorization}
\DeclareAcronym{qos}{short = QoS, long = quality of service}
\DeclareAcronym{cli}{short = CLI, long = command line interface}

\DeclareAcronym{os}{short = OS, long = operating system}
\DeclareAcronym{vm}{short = VM, long = virtual machine}
\DeclareAcronym{vc}{short = VC, long = virtualization container}
\DeclareAcronym{cmd}{short = CMD, long = container management daemon}
\DeclareAcronym{rac}{short = RAC, long = restricted application container}
  
\DeclareAcronym{8021x-s}{short = 802.1X~S, long = 802.1X supplicant}
\DeclareAcronym{8021x-a}{short = 802.1X~A, long = 802.1X authenticator}
\DeclareAcronym{8021x-as}{short = 802.1X~AS, long = 802.1X authentication server}
\DeclareAcronym{8021x-cs}{short = 802.1X~CS, long = 802.1X container supplicant}
\DeclareAcronym{8021x-ca}{short = 802.1X~CA, long = 802.1X container authenticator}
\DeclareAcronym{eapolan}{short = EAPoL, long = EAP-over-LAN}
\DeclareAcronym{eapoudp}{short = EAPoUDP, long = EAP-over-UDP}
\DeclareAcronym{eaporadius}{short = EAPoRADIUS, long = EAP-over-RADIUS}
\DeclareAcronym{eap}{short = EAP, long = Extensible Authentication Protocol}
\DeclareAcronym{radius}{short = RADIUS, long = Remote Authentication Dial In User Service}

\DeclareAcronym{uand}{short = UAND, long = user authentication data}
\DeclareAcronym{cand}{short = CAND, long = container authentication data}
\DeclareAcronym{cazd}{short = CAZD, long = container authorization data}
\DeclareAcronym{vsa}{short = VSA, long = vendor-specific attribute}

\usepackage{tikz}

\usepackage[]{algorithm2e}
\DontPrintSemicolon
\SetAlCapSkip{1em}
\SetKwInOut{Input}{input}
\SetKwInOut{Output}{output}
\SetKwFunction{Print}{print}
\SetKwFunction{AllPaths}{allPaths}
\SetKwFunction{End}{end}

  \newlength{\figsize}
  \newlength{\subfigwidth}
  \newlength{\subfiglabelwidth}
\newcommand\fig[1]{Figure~\ref{fig:#1}}

\newcommand\sect[1]{Section~\ref{sec:#1}}

\begin{document}

  \title{xRAC: Execution and Access Control for Restricted Application Containers on Managed Hosts}
  \author{
    \IEEEauthorblockN{
      Frederik~Hauser,~
      Mark~Schmidt,~
      and Michael~Menth
      \thanks{This work was supported by the bwNET100G+ project which is funded by the Ministry of Science, Research and the Arts Baden-Württemberg (MWK).
        The authors alone are responsible for the content of this paper.}
      }
    \IEEEauthorblockA{
        Chair~of~Communication~Networks,
        University~of~Tuebingen,
        Tuebingen,
        Germany\\
  Email:
        \{%
      frederik.hauser,%
      mark-thomas.schmidt,%
      menth\}@uni-tuebingen.de
    }
}

\maketitle
\pagenumbering{gobble}
  
\begin{abstract}
We propose xRAC to permit users to run special applications on managed hosts and to grant them access to protected network resources.
We use restricted application containers (RACs) for that purpose.
A RAC is a virtualization container with only a selected set of applications.
Authentication verifies the RAC user's identity and the integrity of the RAC image.
If the user is permitted to use the RAC on a managed host, launching the RAC is authorized and access to protected network resources may be given, e.g., to internal networks, servers, or the Internet.
xRAC simplifies traffic control as the traffic of a RAC has a unique IPv6 address so that it can be easily identified in the network.
The architecture of xRAC reuses standard technologies, protocols, and infrastructure.
Those are the Docker virtualization platform and 802.1X including EAP-over-UDP and RADIUS.
Thus, xRAC improves network security without modifying core parts of applications, hosts, and infrastructure.
In this paper, we review the technological background of xRAC, explain its architecture, discuss selected use cases, and investigate on the performance.
To demonstrate the feasibility of xRAC, we implement it based on standard components with only a few modifications. 
Finally, we validate xRAC through experiments.
\end{abstract} 
  
\IEEEpeerreviewmaketitle
 
\graphicspath{{figures/}}

\section{Introduction}
\label{sec:introduction}

In this paper we consider the problem of permitting users to run special applications on managed hosts and to grant them access to protected network resources.
This is an important challenge in practice as applications communicate with multiple peers and have multiple, possibly a priori unknown flows characterized by 5-tuples.
Moreover, common packet filters such as firewalls or deep packet inspectors lack knowledge of allowed flows, and identifying traffic from specific applications becomes more difficult with traffic encryption using TLS.

We address this challenge by running applications in containers as so-called \acp{rac} on managed hosts.
\acp{rac} provide selected sets of applications including their dependencies and configuration.
The managed host gives users only limited freedom, e.g., they can download and launch \acp{rac}.
We propose \ac{aa} for \acp{rac} so that their execution is restricted to authorized users.
That means, the user identity, the integrity of the \ac{rac}'s image, and the permission of the user to execute the \ac{rac} are verified before a \ac{rac} is launched.
We suggest to extend this authorization also to protected network resources required by the \ac{rac}, e.g., to internal networks, servers, or Internet access.
That means, appropriate network control elements, e.g., firewalls or SDN controllers, may be informed about authorized \acp{rac} and their needs.
The authorized traffic can be identified by the \ac{rac}'s IPv6 address.
We call this concept xRAC as it controls execution and access for \acp{rac}.

We mention a few use cases that may benefit from xRAC.
\acp{rac} may allow users to execute only up-to-date \ac{rac} images.
Execution of a \ac{rac} may be allowed only to special users or user groups, e.g., to enforce license restrictions.
Only selected users in a high-security area may get access to the Internet through a \ac{rac}-based browser which is isolated from the remaining infrastructure.
Only selected users may be able to execute administration software with access to servers providing confidential material.
\acp{rac} may be used for applications with increased \ac{qos} requirements, e.g., voice-over-IP, video conferences, or games.
Their traffic may be preferentially treated by network elements.

To facilitate deployment of xRAC, we reuse and adapt standard technologies, protocols, and infrastructure.
We leverage the Docker platform to create and deploy containerized applications.
When the \ac{cmd} is requested to launch a \ac{rac}, it first issues an \ac{aa} request and launches the \ac{rac} only after successful \ac{aa}.
We adopt 802.1X for \ac{aa} purposes and adapt its components 802.1X supplicant (802.1X~S), 802.1X authenticator (802.1X~A), and 802.1X authentication server (802.1X~AS).
The \ac{cmd} interfaces with an \ac{8021x-cs} on the host, the \ac{8021x-cs} with a \ac{8021x-ca}, and the \ac{8021x-ca} with an \ac{8021x-as}.
The \ac{8021x-as} holds \ac{rac}-specific \ac{aa} data, performs authentication, and returns authorization data to the \ac{8021x-ca}.
The \ac{8021x-ca} interfaces with network control elements and configures access to network resources depending on authorization data.
It also forwards the authorization data to the \ac{8021x-cs}.
EAP-over-UDP and RADIUS are utilized for communication.
To demonstrate the feasibility of xRAC, we provide a prototype based on existing 802.1X components, implement the \ac{8021x-cs} as plugin for the Docker authorization framework and the \ac{8021x-ca} as part of an SDN controller.
We use a RADIUS server for \ac{8021x-as} and extend its data structures to store \ac{aa} data for users and \acs{rac}.
We use this prototype to experimentally validate xRAC in a testbed.

The rest of the paper is structured as follows.
In \sect{container}, we review technical background and related work on container virtualization.
In \sect{8021x}, we review technical background and related work on 802.1X and \ac{aa} for applications.
In \sect{concept}, we present the architecture of xRAC in detail.
\sect{use-cases} discusses use cases along benefits and limitations of xRAC.
\sect{implementation} describes the prototypical implementation of xRAC which is used for its experimental validation in \sect{functional-validation}.
In \sect{performance-considerations}, we briefly investigate on performance considerations of xRAC.
\sect{conclusion} concludes this work.
\section{Container Virtualization: Technical Background and Related Work}
\label{sec:container}

We first introduce the concept of container virtualization, describe advantages and give an overview of Docker as a widespread implementation.
Then, we review container security platforms and containers for GUI applications.

\subsection{Overview}
Containers implement virtualization on the \ac{os} level.
They provide virtualized \ac{os} environments that are isolated with regard to hardware resources and security.
They share the \ac{os} kernel and may include binaries and libraries that are required to run one or several enclosed applications.
In xRAC, we enclose only one special application with its dependencies and configuration in a container.
Containers run on top of a container runtime and are managed by a \acf{cmd} that creates, starts and suspends containers.
Examples for container platforms are Docker \cite{docker}, Kubernetes \cite{kubernetes}, BSD jails \cite{bsd-jails}, and Solaris containers \cite{solaris-containers}.

\subsection{Advantages}
\label{sec:container-advantages}
Virtualization facilitates efficient and flexible usage of hardware resources, improves security through isolation, and provides fault-safety and scalability through simple migration processes.
Containers in particular have the following additional advantages.
Due to the shared \ac{os}, containers require less CPU, memory, and hard disk resources.
Container images are much smaller than virtual machines, which simplifies distribution among many recipients.
Containers simplify application deployment.
Instead of providing support for complex combinations of applications, dependencies, and user configurations, administrators just deploy containers that are tested prior to release.
Containers have no bootup times, which makes them especially suitable for short-lived applications.

\subsection{Docker Container Virtualization Platform}
\begin{figure}[htp]
    \centering
    \includegraphics[width=.95\linewidth]{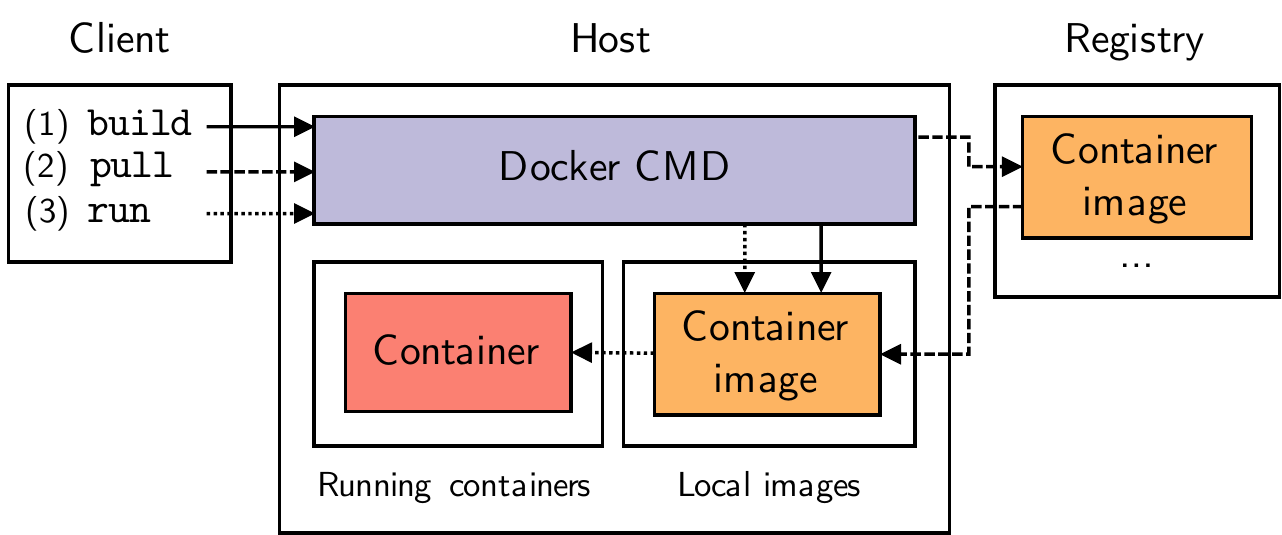}
    \caption{The Docker architecture consists of the Docker client, the Docker \ac{cmd}, and the Docker registry \cite{docker-getstarted}.}
    \label{fig:docker-architecture}
\end{figure}

Docker \cite{docker} is one of the most popular container platforms today.
\fig{docker-architecture} provides a simplified overview of the Docker platform including its operations.
A host with a common \ac{os} runs the Docker \ac{cmd} that generates and manages containers and container images.
Container images are write-protected templates that include applications with their dependencies.
Containers are runtime instances that extend the write-protected container images by a writeable layer.
Therefore, multiple container instances may share a common image.
This introduces scalability with low hard disk requirements.
The Docker \ac{cmd} is controlled by the Docker client via a REST interface.
The Docker client can be located on the host running the Docker \ac{cmd} or on a remote host.
The Docker \ac{cli} is an example for user control through \ac{cli} calls.
The Docker \ac{cmd} may connect to Docker registries that allow users to upload (push) or download (pull) container images.
Those registries are either private or publicly available.
Docker Hub \cite{docker-hub} with more than 100.000 container images is an example for the latter.
Common operations are build (1), pull (2), and run (3).
With build, users may create individual container images.
With pull, users may download container images from a repository to become part of the set of local images.
With run, container images from the set of local images can be executed on the host system.

\vspace{-0.25cm}
\vspace{-0.15cm}

Docker uses several functions of the \ac{os} kernel \cite{cgroups1, cgroups2, ns1, ns2} to provide isolation and resource emulation.
It supports storage drivers, e.g., AuFS \cite{aufs}, OverlayFS \cite{overlayfs}, and ZFS \cite{zfs}, to enable file system stacking.
The container format and runtime environment of Docker were adopted by the Open Container Initiative \cite{oci} as open industry standard.

\subsection{Container Security Platforms}
\label{sec:container-security-platforms}
Container security platforms extend the \ac{cmd} by security functions.
Twistlock \cite{twistlock, twistlock-datasheet} and the Aqua Container Security Platform \cite{aqua} provide a runtime engine based on machine learning mechanisms to permanently monitor containers for detecting fraudulent behavior and special network firewalls to filter container traffic.
The Sysdig Secure \cite{sysdig-secure} platform allows the formulation of service-aware policies, i.e., policies that are based on applications, containers, hosts, or network activities.
The platform provides alerts and actions based on policy violations, an event history, and incident captures.
The Atomicorp Secure Docker Kernel \cite{atomicorp-docker} is a hardened Linux kernel that provides security-related features such as break-out protection, memory corruption protection, or prevention of direct userland access by the kernel.
All platforms focus on monitoring and controlling potentially untrustworthy containers that are executed on a shared container runtime.
Features for \ac{aa} for users, containers, and their network flows are not part of those platforms.

The Docker Authorization Framework \cite{docker-authz} is part of Docker since Version 1.10 \cite{docker-authz-announcement}. 
It extends the Docker \ac{cmd} by a REST interface to external authorization plugins.
Requests from the Docker \ac{cmd}, e.g., to start a container, are forwarded to an authorization plugin that implements mechanisms to decide whether to allow or deny the request.
The Docker Authorization Framework does not implement security functions but provides a base for implementing such security concepts.
xRAC leverages this framework.

\subsection{Containers for GUI Applications}
Containers typically deploy applications or services without \acp{gui} that run in the cloud or on data center infrastructures.
Examples are containers that enclose web applications with their requirements, e.g., an nginx webserver with a PHP runtime and a MySQL database server.
The idea of leveraging Docker containers to deploy desktop applications with a \ac{gui} was first presented in \cite{docker-desktop}.
The author proposes to mount X11 sockets for \ac{gui} presentation and hardware devices of the host system, e.g., an audio card or a web camera, to the Docker container.
Thereby, even more complex \ac{gui} applications such as the Chrome web browser, the Spotify music player, or the Skype video chat application can be run in containers and be deployed as container images.
Today, many Docker container images with \ac{gui} applications can be downloaded from Docker Hub.
\section{802.1X: Technical Background and Related Work}
\label{sec:8021x}

We give an overview of 802.1X and explain how it supports \ac{aa}.
We present EAPoUDP which is an alternative protocol to carry \ac{aa} data in 802.1X.
We summarize how \ac{aa} for applications is currently performed in practice and review another research work that provides \ac{aa} for applications based on 802.1X.

\subsection{Overview}
IEEE 802.1X \cite{8021X_2001, 8021X_2004, 8021X_2010} introduces port-based network access control in wired Ethernet networks.
However, it is mainly known from wireless 802.11 networks today.
An example is Eduroam \cite{eduroam}, a federation of wireless university campus networks.
Participants can connect to the Internet no matter if located at their home institution or at a foreign university campus, e.g., while attending a conference.

\begin{figure}[htp]
    \centering
    \includegraphics[width=.95\linewidth]{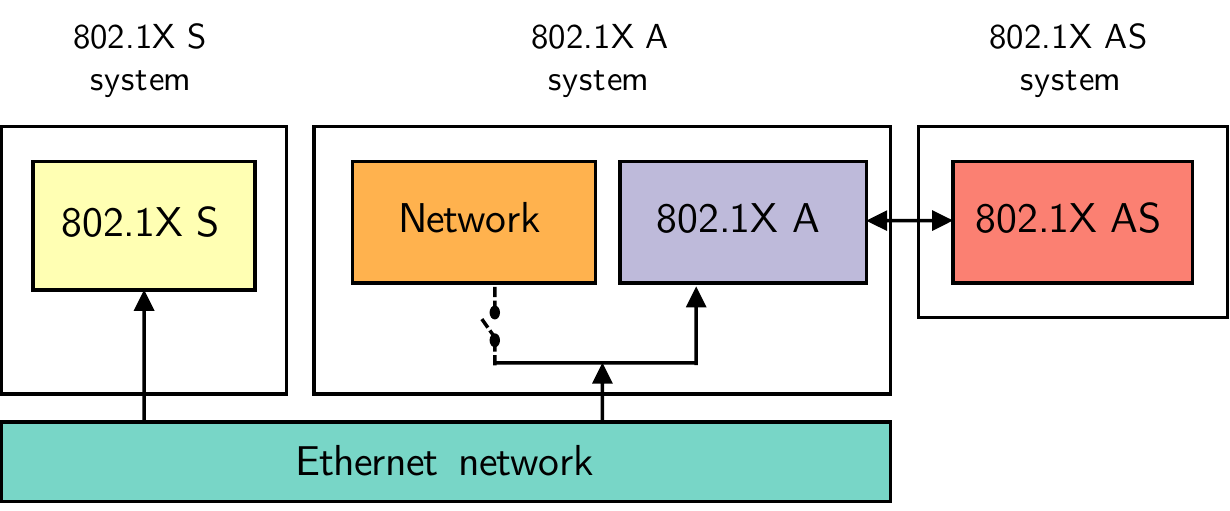}
    \caption{Port-based authorization model of 802.1X \cite{8021X_2004}.}
    \label{fig:8021X_Roles}
\end{figure}

\fig{8021X_Roles} depicts the three components of 802.1X and the principle of port-based network access control.
The supplicant system is a network host that contains the \ac{8021x-s}.
The authenticator system contains the \ac{8021x-a} and controls network access of network hosts.
Examples are access switches that connect network hosts to the main network.
Prior to authorization, the supplicant system can reach the \ac{8021x-a} but not the network.
The \ac{8021x-as} is an authentication, authorization, and accounting server.
It stores authentication data to verify user identities and authorization data to grant permission to the network.
It authenticates the \ac{8021x-s} and delivers authorization information to the \ac{8021x-a}.

\vspace{-0.15cm}

\subsection{\ac{aa} with 802.1X}
802.1X leverages the \ac{eap} \cite{rfc3748} and the \ac{radius} \cite{rfc2865} to exchange \ac{aa} data.
Both provide a fixed request and response scheme to exchange \ac{aa} data.
The Diameter protocol \cite{rfc6733} is a less widespread alternative.
Authentication data is transmitted in Ethernet frames as \ac{eapolan} encapsulation between the \ac{8021x-s} and \ac{8021x-a} and as \ac{eaporadius} between the \ac{8021x-a} and \ac{8021x-as}.
\fig{eapoudp} depicts the packet structure of \ac{eapolan}.
Authorization data is transmitted in \ac{radius} frames between the \ac{8021x-as} and \ac{8021x-a}.

\begin{figure}[htp]
    \centering
    \includegraphics[width=.95\linewidth]{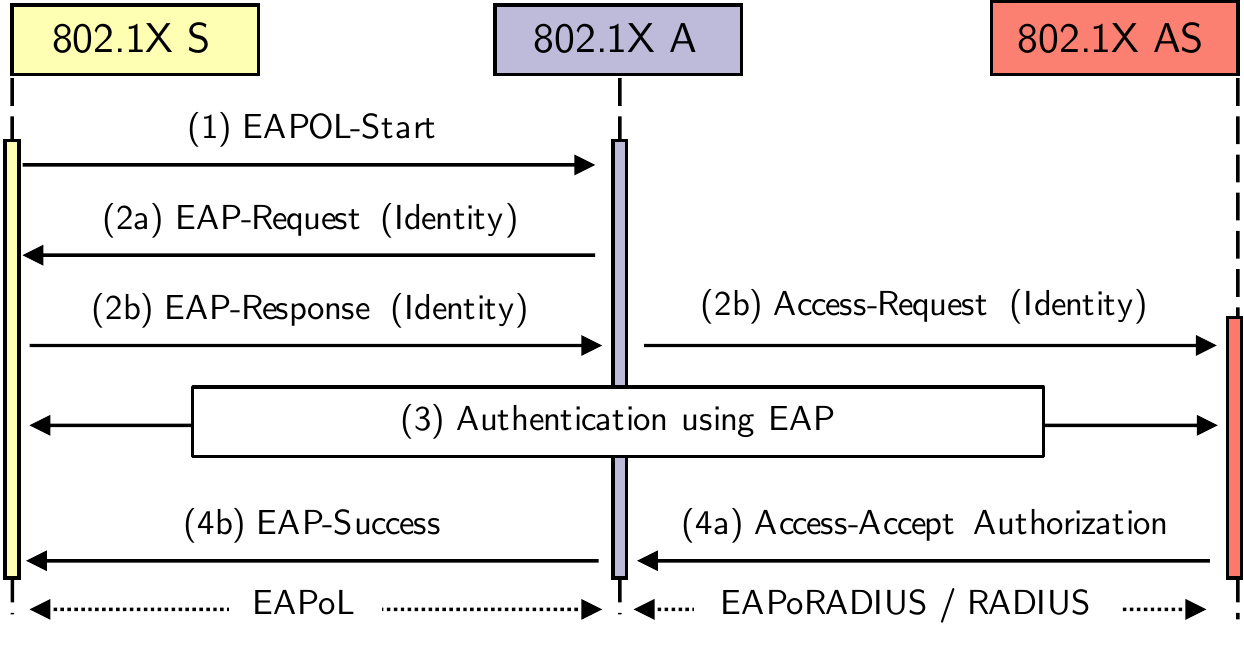}
    \caption{Communication example of 802.1X based \ac{aa}.}
    \label{fig:8021X_Procedure}
\end{figure}

We explain the details of 802.1X with the four-step process of \ac{aa} as depicted in \fig{8021X_Procedure}.
In the first step (1), the \ac{8021x-s} initializes authentication by sending an \mbox{EAPOL-Start} message to the \ac{8021x-a}.
In the second step, the \ac{8021x-a} requests the identity from the \ac{8021x-s} (2a) and forwards it to the \ac{8021x-as} (2b). 
\ac{radius} supports large domains that consist of many hierarchically organized \ac{radius} servers. 
Each identity is associated with a domain and known by the \ac{radius} server of that domain so that \ac{aa} attempts can be forwarded in \ac{radius} infrastructures.
In the third step (3), authentication is performed between the \ac{8021x-s} and \ac{8021x-as}. 
The authenticator decapsulates \ac{eap} packets from \ac{eapolan} frames and reencapsulates them as \ac{eaporadius} frames and vice versa. 
The flexible message structure of \ac{eap} allows the use of different authentication procedures.
Simple approaches carry plain-text identity information or simple MD5-hashed passwords, but more secure authentication procedures like EAP Tunneled TLS \cite{rfc5281} and EAP-TLS \cite{rfc5216} are also supported.
The authenticator only relays \ac{eap} messages in pass-through manner.
Therefore, new \ac{eap} types only need to be implemented on the \ac{8021x-s} and \ac{8021x-as} but not on the \ac{8021x-a}.
In the fourth step, the \ac{radius} server may return authorization data after successful authentication to the \ac{8021x-a} (4a).
It can be coarse-granular, e.g., a binary access decision whether the supplicant system gets access or no access, or fine-granular, e.g., VLAN tags \cite{rfc4675} to be set for prospective user traffic or filter rules \cite{rfc4849} that are applied by the authenticator.
The authenticator applies the authorization data on the particular physical port of the switch, e.g., it sets a VLAN tag.
Afterwards, the authenticator confirms successful \ac{aa} to the supplicant with an EAP-Success message (4b).

\vspace{-0.15cm}

\subsection{\acf{eapoudp}}
\begin{figure}[htp]
    \centering
    \includegraphics[width=.9\linewidth]{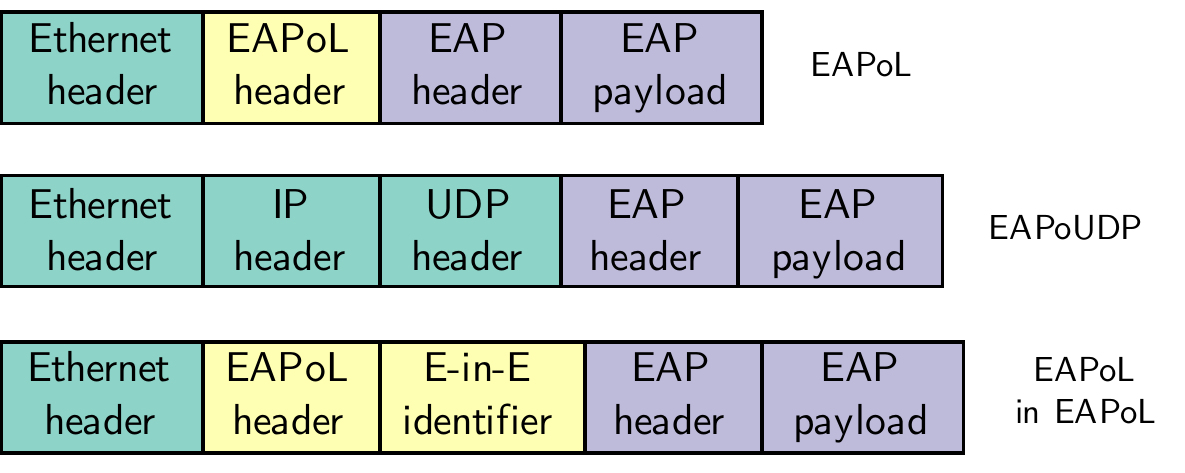}
    \caption{\ac{eapoudp} (a) in comparison to EAPoL (b). \ac{eapoudp} uses UDP as transport protocol, EAPoL leverage Ethernet frames along an EAPoL header.}
    \label{fig:eapoudp}
\end{figure}

\ac{eapoudp} is a variation of \ac{eap} that allows transmission of EAP data over UDP and IP.
\fig{eapoudp} depicts its packet structure in comparison to EAPoL.
In contrast to \ac{eapolan}, \ac{eapoudp} can be used to authenticate multiple applications that run on a network host.
Also, UDP packets can be transmitted over any link layer technology or even routed within multi-domain networks.
\Ac{eapoudp} was introduced as Internet draft \cite{eapoudp-draft} that expired without standardization in the PANA working group of the IETF in 2002.
Cisco leveraged \ac{eapoudp} in its Trust Agent \cite{cisco-trust-agent} tool that runs on network hosts and interacts with Cisco NAD, a prorietary network control system.
The Trust Agent collects host system information, interfaces host software, and delivers notifications to network hosts within \ac{eapoudp} frames.
xRAC leverages \ac{eapoudp} for frontend authentication.

\subsection{AA for Applications in Practice}
802.1X focuses on port-based access control for network hosts.
In practice, \ac{aa} for applications is implemented as part of the application or with the help of Kerberos.

Most network applications implement some form of \ac{aa} mechanism. 
Examples are login forms in the launch window where users are required to enter valid credentials to start using the applicaton.
Other examples are client certificates that are used in conjunction with TLS and a public key infrastructure.
However, \ac{aa} functionalities that are part of the application have an impact that is limited to the client and server side of the network application.
Neither the start of the application nor the network infrastructure in between can be controlled.

Kerberos~\cite{kerberos, rfc4120} is a network authentication protocol that provides mutual authentication for clients and servers over an insecure network.
Clients are entire hosts, users, or applications; servers represent hosts that offer particular network applications.
Kerberos adapts user tickets for authentication for various network applications.
Kerberos needs to be implemented by applications on client and server side, which prevents its application for legacy applications that cannot be modified.
Again, neither the start of kerberized applications nor the network infrastructure in between can be controlled.

\subsection{AA for Applications with FlowNAC}
FlowNAC \cite{flownac} introduces a fine-granular SDN network access control system that adapts 802.1X for \ac{aa} of applications on network hosts.
To enable multiple \ac{aa} for different applications on a network host, the authors introduce EAPoL-over-EAPoLAN encapsulations.
As depicted in \fig{eapoudp}, FlowNAC introduces another variation of EAPoL.
An EAPoL-in-EAPoL packet field identifies up to 64K different EAP processes that are transmitted as encapsulated EAP payloads.
However, this deviation from legacy 802.1X requires major changes of the \ac{8021x-s} and \ac{8021x-a}.
The \ac{8021x-s} is part of an \ac{os}'s kernel, the \ac{8021x-a} is part of network switches so only open source \acp{os} and firmwares allow modifications.
Nevertheless, it is difficult to carry on the modifications in new versions of the \ac{os}'s kernel or firmware image. 
The authors rely on \ac{eapolan}, i.e., \ac{aa} data transfer is limited to the Ethernet link.
\ac{eapoudp} would solve those shortcomings but was not considered in the work.
Unlike xRAC, FlowNAC neither introduces IP addresses for applications nor restricts the start of applications by \ac{aa}. 
\section{xRAC Architecture}
\label{sec:concept}

In this section, we first explain \acp{rac} and give an overview of xRAC.
Then, we explain the operation of its three control components in details.

\subsection{Restricted Application Containers (RACs)}
\begin{figure}[htp]
    \centering
    \includegraphics[width=.98\linewidth]{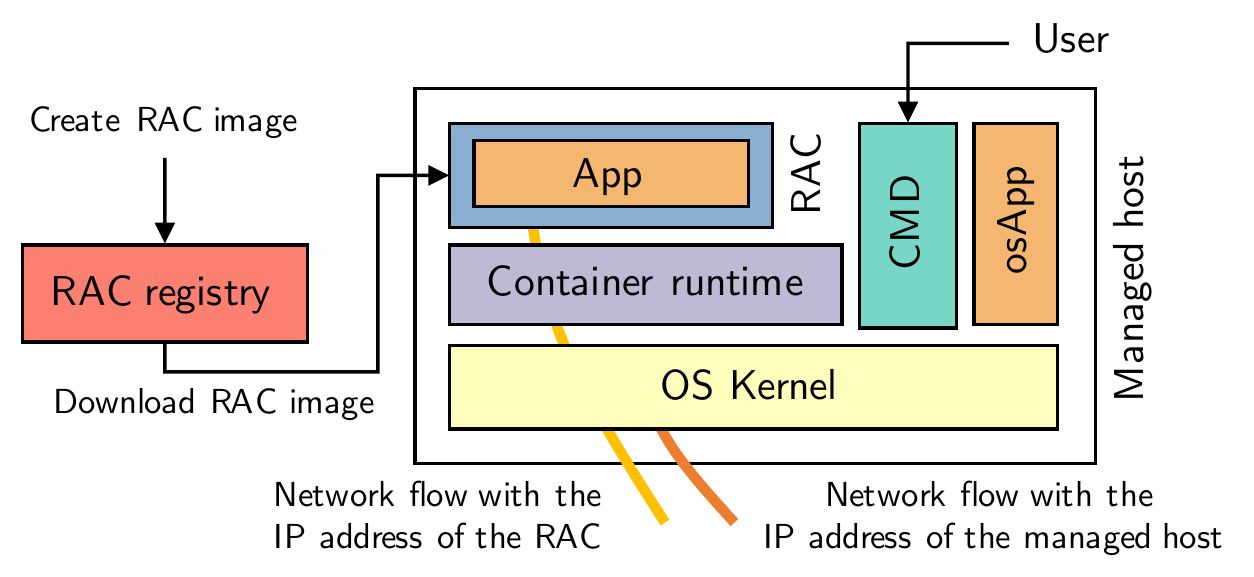}
    \caption{The managed host executes a \ac{rac} and a \ac{os} application (osApp).}
    \label{fig:capps}
    \vspace{-0.3cm}
\end{figure}

\acp{rac} are executable container images that enclose a single application, its dependencies, and configuration data such as program settings or software licensing information.
As depicted in \fig{capps}, \acp{rac} are executed on a container runtime in parallel to \ac{os}-native applications.
The \ac{cmd} controls the execution of \acp{rac} and provides an interface for users to create, delete, start, or stop \acp{rac}.
Each \ac{rac} has a unique IPv6 address so that its traffic can be easily identified in the network.
\ac{rac} images are created by network administrators and deployed via \ac{rac} registries.
They are either downloaded from those registries by users or automatically synchronized to managed hosts.

\vspace{-0.3cm}

\subsection{Overview of xRAC}
\begin{figure}[htp]
    \centering
    \includegraphics[width=.98\linewidth]{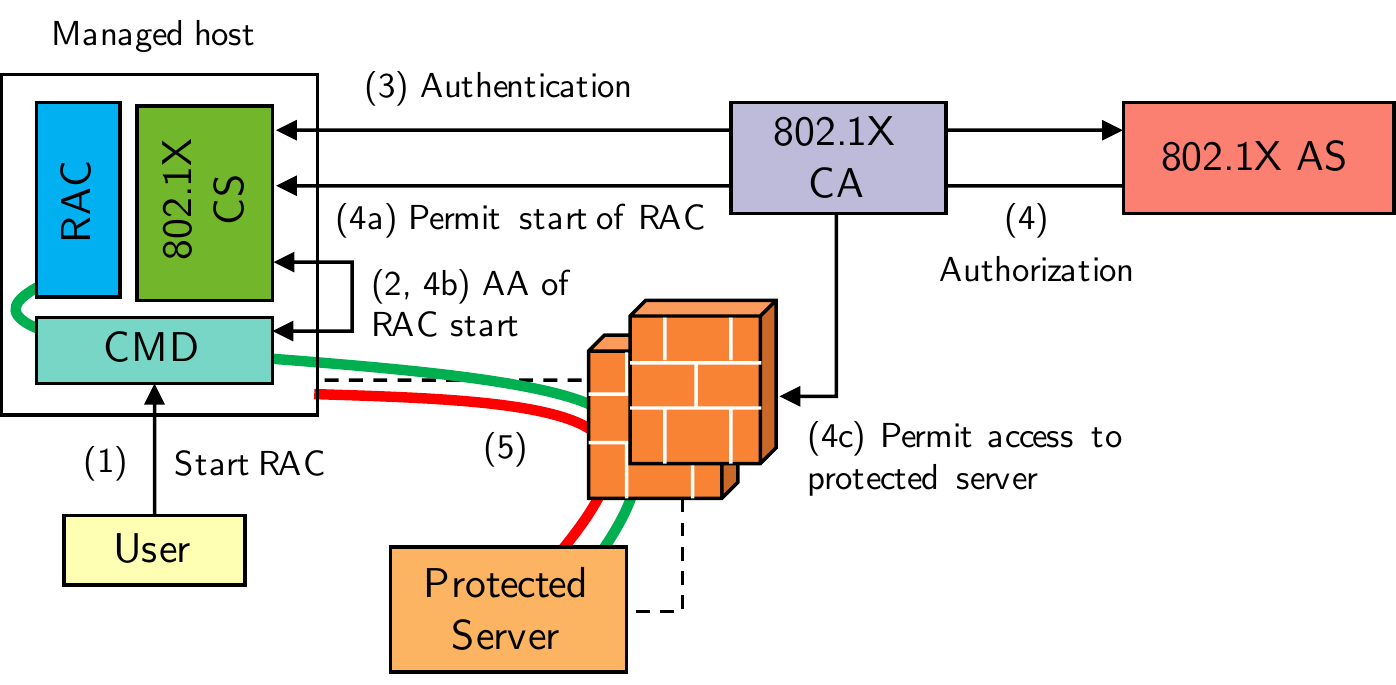}
    \caption{Adaption of 802.1X for \ac{aa} of \acp{rac}. The \ac{8021x-cs} authenticates the \ac{rac} using the \ac{8021x-as}. The \ac{8021x-ca} receives authorization data and forwards them to the \ac{8021x-cs} and to a firewall that secures a protected server from unauthorized traffic.}
    \label{fig:concept-makro}
\end{figure}

xRAC provides execution and access control for \acp{rac} on managed hosts.
A \ac{rac} needs to be authenticated and authorized before launching.
\fig{concept-makro} depicts the \ac{aa} process for \acp{rac} with 802.1X.
First, a user attempts to start a \ac{rac} via the \ac{cmd} (1), and the \ac{cmd} requests the \ac{8021x-cs} for \ac{aa} (2).
After successful authentication (3), the \ac{8021x-as} responds with authorization data via the \ac{8021x-ca} (4) to the \ac{8021x-cs} (4a).
The \ac{8021x-cs} notifies the \ac{cmd} to launch the \ac{rac} (4b).
In addition, the \ac{8021x-ca} informs network control elements about the authorized \ac{rac}.
In our example, it configures the firewall to permit access to the proteced server (4c).
Other examples are SDN controllers that program SDN switches.
Now, the authorized \ac{rac} but not the managed host or other \acp{rac} can communicate with the protected server (5).

\vspace{-0.2cm}

\Ac{aa} for \acp{rac} introduces two additional advantages to common application deployment and network security.
First, \ac{aa} of \acp{rac} restricts \ac{rac} launches on managed hosts to predefined \ac{rac} images and permitted users.
This allows network operators to ensure that only up-to-date and unmodified \ac{rac} images can be launched.
This improves computer and network security as only valid \ac{rac} images can be executed on the managed hosts.
In addition, network operators may deploy \ac{rac} images to managed hosts in advance, e.g., by synchronizing their set of \ac{rac} images with an internal \ac{rac} repository in the background.
Users have all available \ac{rac} images on managed hosts but are only able to start them if they become authorized after \ac{aa}.
Last, each \ac{rac} has a globally unique IPv6 address that can be used to identify and steer all traffic originating from a particular \ac{rac}.
\ac{rac} authorization data on the \ac{8021x-as} includes information on how the \ac{rac}'s traffic should be steered by network elements that can be applied by network control elements.

\subsection{802.1X Authentication Server (802.1X AS)}
\begin{figure}[htp]
    \centering
    \includegraphics[width=.98\linewidth]{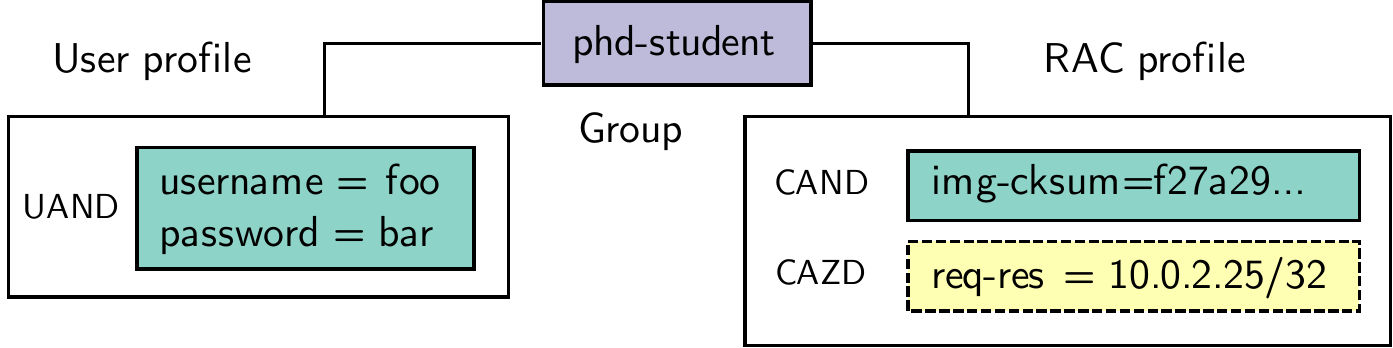}
    \caption{The \ac{aa} data model for \acp{rac} consists of user profiles, \ac{rac} profiles, and group mappings. User profiles include \ac{uand}, \ac{rac} profiles contain \ac{cand} for authentication and \ac{cazd} for authorization.}
    \label{fig:radius-vc}
    \vspace{-0.3cm}
\end{figure}

The authentication request from the \ac{8021x-cs} to the \ac{8021x-as} contains user and container authentication data (UAND, CAND).
The 802.1X AS authenticates the user and verifies the integrity of the RAC image.
If the RAC image is valid, and if the user is authenticated, and if the user has permissions to run the RAC, the 802.1X AS responds to the 802.1X CA with authorization data.
To perform that decision, the 802.1X AS requires a new data model which is depicted in \fig{radius-vc}.
It consists of user profiles (1), \ac{rac} profiles (2), and groups (3) that define whether a particular user is permitted to run a particular \ac{rac}.
User profiles (1) contain \ac{uand} that is used to authenticate the user.
Examples are user credentials, e.g., user names and passwords.
\ac{rac} profiles (2) contain \ac{cand} and \ac{cazd}.
The first is used to verify the integrity of the \ac{rac} through calculating the cryptographic hashing function over the \ac{rac} image.
\ac{cazd} include all permissions of a \ac{rac}, i.e., to be started by the requesting user and to utilize network resources.
In the depicted example, the \ac{rac} is allowed to access a specified network resource.
The \ac{aa} data for the described model is stored on the \ac{8021x-as}.
The data model is an example that can be easily extended to support other requirements.

\vspace{-0.2cm}

\subsection{802.1X Container Supplicant (802.1X CS)}
The \ac{8021x-cs} authenticates \acp{rac} with the \ac{8021x-as} via the \ac{8021x-ca}.
It transmits \ac{uand} and \ac{cand} to the \ac{8021x-as} and receives \ac{cazd} from the \ac{8021x-ca}.

\fig{8021xcs} illustrates the process of \ac{aa} from the perspective of the \ac{8021x-cs}.
It runs on the managed host, interfaces the \ac{cmd}, and is configured with the IP address or URL of the \ac{8021x-ca} so that it can initiate \ac{aa}.
In (1), the user requests the \ac{cmd} to start a particular \ac{rac} on the managed host.
The request includes \ac{uand}.
The \ac{cmd} requests the \ac{8021x-cs} to permit the user's demand (2).
The request includes \ac{uand} and \ac{cand} collected by the \ac{cmd}.
The \ac{8021x-cs} initiates \ac{aa} by establishing an \ac{eapoudp} session with the \ac{8021x-ca}.
Afterwards, it performs authentication with the \ac{8021x-as} via the \ac{8021x-ca} (3).
Backend authentication is performed via \ac{eaporadius} while frontend authenticated is performed via \ac{eapoudp} as in legacy 802.1X.
In case of successful authentication, the 802.1X CS receives CAZD from the 802.1X CA (4).
Then, the 802.1X CS permits the CMD to start the RAC (5).

\subsection{802.1X Container Authenticator (802.1X CA)}
The 802.1X CA relays AA data between the 802.1X CS and the 802.1X AS.
Moreover, it informs network control elements about authorized RACs.

\begin{figure}[htp] 
    \centering 
    \begin{subfigure}[b]{.95\linewidth} 
        \centering 
        \includegraphics[width=\textwidth]{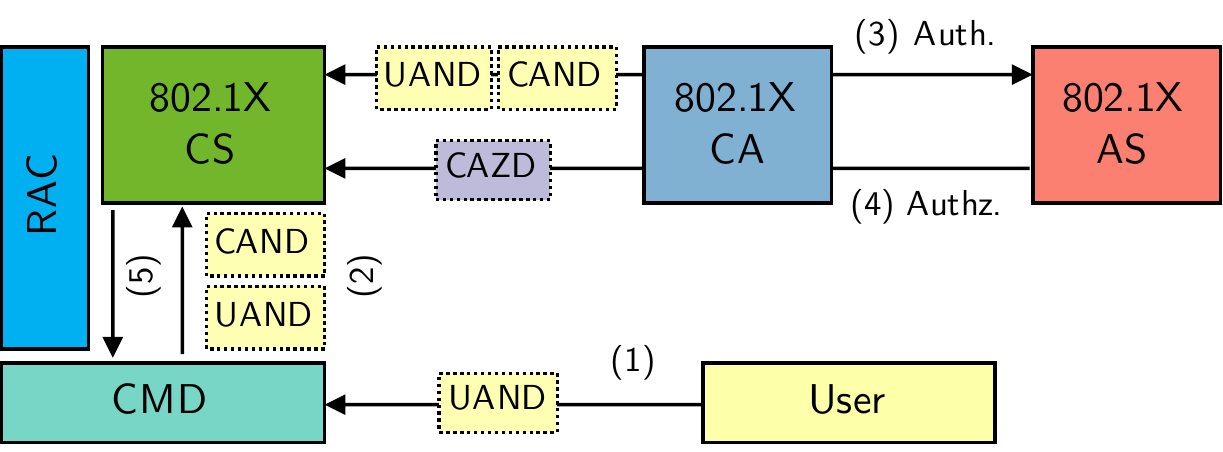} 
        \caption{\footnotesize{\ac{8021x-cs} receives \ac{uand} and \ac{cand} from the \ac{cmd}. It then performs frontend authentication via \ac{eapoudp} and backend authentication via \ac{eaporadius} with the \ac{8021x-as}. In case of successful authentication, the \ac{8021x-ca} and the \ac{8021x-cs} receive \ac{cazd}.}}
        \label{fig:8021xcs} 
    \end{subfigure}
    \begin{subfigure}[b]{.95\linewidth} 
        \centering 
        \includegraphics[width=\textwidth]{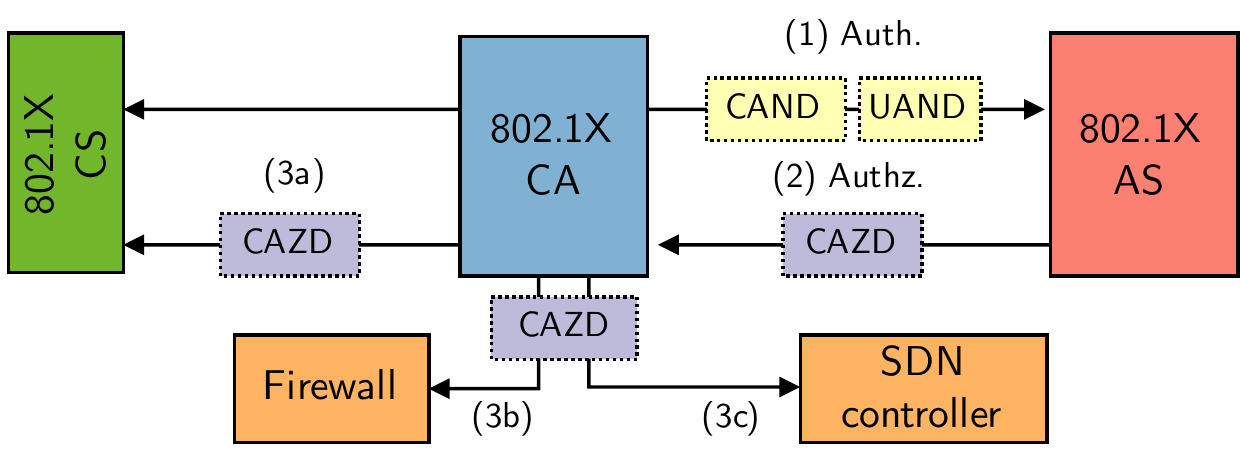} 
        \caption{\footnotesize{The \ac{8021x-ca} performs frontend authentication with the \ac{8021x-cs} and backend authentication with the \ac{8021x-as}. In case of successful authentication, it receives \ac{cazd} from the \ac{8021x-as} that is forwarded to the \ac{8021x-cs} and other network components.}}
        \label{fig:container-authenticator} 
    \end{subfigure}
    \caption{\ac{aa} from the perspective of the \ac{8021x-cs} and the \ac{8021x-ca}.} 
\end{figure}

In step (1) of \fig{container-authenticator}, authentication data are transported over EAP between 802.1X CS and 802.1X AS.
Between 802.1X CS and 802.1X CA, the EAP data are carried over UDP (EAPoUDP) and between 802.1X CA and 802.1X AS, they are carried over RADIUS. 
Thus, one task of the 802.1X CA is to modify the tunnel for EAP data.
Moreover, after successful authentication the 802.1X AS returns CAZD over RADIUS to the 802.1X CA (2) which then informs the 802.1X CS about successful authorization (3a).
While the conventional 802.1X A just opens ports on a switch for authorized devices, the 802.1X CA may also inform other network control elements about authorized RACs.
Those may be ports on a switch, firewalls (3b), or SDN controllers (3c).
The firewall is then programmed to pass through all outbound traffic with the RAC's IP address and the SDN controller instructs SDN switches to forward all traffic with the RAC's IP address appropriately.
More specific flow descriptors are not needed.

\section{Use Cases and Discussion}
\label{sec:use-cases}

We discuss two exemplary use cases of xRAC and discuss its benefits
and limitations.

\subsection{Use Case I: Web Browsers in High-Security Areas}
Research departments, state departments, or clinics dealing with highly sensitive data isolate their internal networks from the Internet.
However, web browsers are still required for online research activities.
We propose to deploy web browsers as \acp{rac} on managed hosts.
The isolation of \acp{rac} prevents that malicious users may misuse the Internet access to leak internals or contaminate the internal network through infectious downloads, e.g., PDF documents that include a trojan or virus.
The network flow control of \acp{rac} ensures, that only the web browser can reach the Internet.
If the \ac{rac}'s traffic is encrypted, e.g., DNS queries and web site data, network control elements can still perform packet filtering based on the IP addresses of the \ac{rac}.

\subsection{Use Case II: Confidential Data Access}
Applications dealing with confidential data, e.g., research activities, medical documentation, or law enforcement, often access servers with confidential data.
If such applications are deployed as RACs on managed hosts, only legitimate users have access to those servers.
The isolation feature of RACs prevents remote hackers from attacking the server.
Normally, they get system access through gateways provided by viruses or trojans received as browser downloads or e-mail attachments which is not possible with \acp{rac}.
Furthermore, malicious users of legitimate applications may use hacker tools to gain access to the server and to leak information from it, which is not possible in the digital domain with an isolated application.
The isolation of \acp{rac} prevent malicious users or applications from attacking the server.
The network flow control ensures that the server can be reached only by legitimate \acp{rac} and users but not by other \acp{rac} or the managed host itself.

\subsection{Benefits of xRAC}
xRAC inherits the advantages generally known from virtualization
and container virtualization that we have discussed in \sect{container-advantages}. In
addition, xRAC can guarantee that only valid containers are
launched on managed hosts and that they can be used only by
legitimate users. Thus, xRAC performs AA for applications without the need to
modify them, which is a particular benefit for legacy applications. Moreover, the 802.1X CA can configure network
control elements such that authorized RACs have access to protected
network resources. RACs facilitate this control as all traffic of a
RAC is identified by a single IPv6 address. This is a particular
benefit as in today's networks there is no information about
legitimate flows, many application flows may have the same IP
address, and applications may even be invisible due to encryption
using TLS. Thus, steering traffic from legitimate or trusted
applications is a tough problem for which xRAC provides a
solution. xRAC is flexible as it implements software-defined network control
by interacting with other network control elements. In particular,
it does not depend on and is not limited to specific technologies.

\subsection{Limitations of xRAC}
xRAC requires a managed infrastructure where the managed host, its
CMD, and the 802.1X components, i.e., 802.1X CS, 802.1X CA, and 802.1X AS, are trusted.
Encapsulating applications in RACs complicates access to shared
resources so that xRAC may be cumbersome or infeasible for some use cases.
 
\section{Prototypical Implementation}
\label{sec:implementation}

In the following, we describe a prototypical implementation of xRAC.
We overview the testbed environment and describe all components in detail.

\subsection{Testbed Environment}
\begin{figure}[htp]
    \centering
    \includegraphics[width=.95\linewidth]{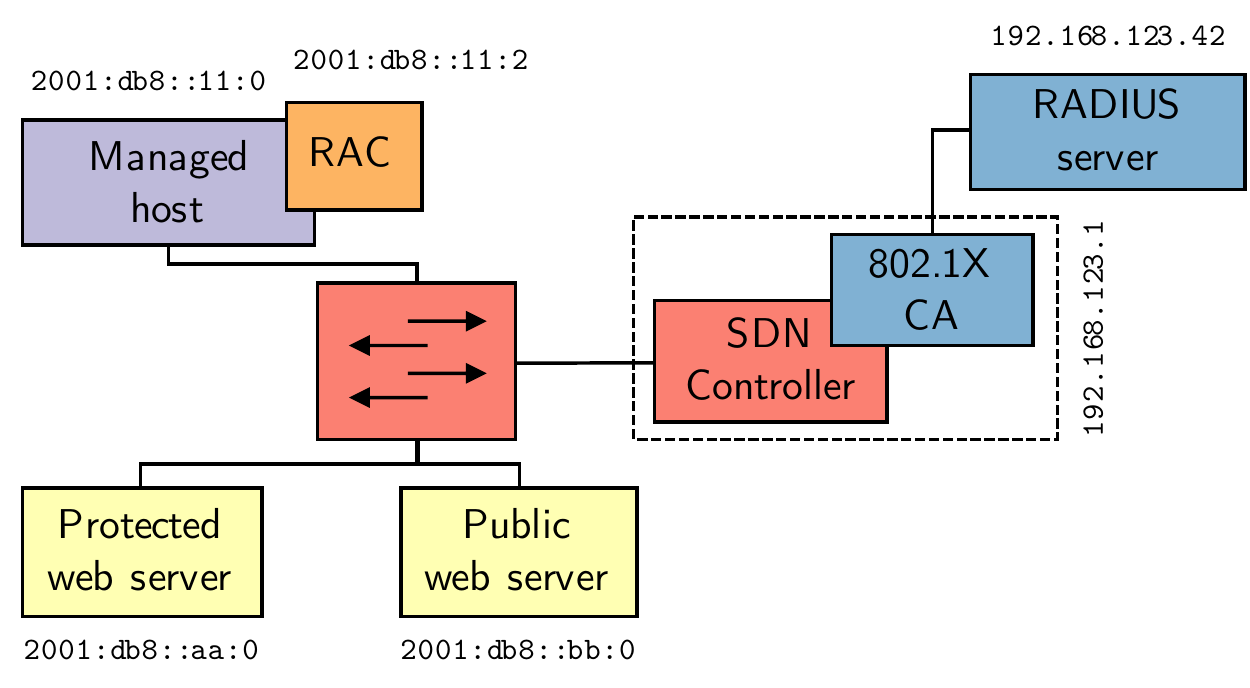}
    \caption{Testbed environment.}
    \label{fig:implementation-overview}
\end{figure}

\fig{implementation-overview} depicts the testbed environment.
The managed host executes \acp{rac}.
The SDN switch connects the managed host, the protected web server, the public web server and is controlled by an SDN controller.
The SDN controller runs the \ac{8021x-ca} as SDN application that communicates with an \ac{8021x-as}.

\vspace{-0.2cm}
\vspace{-0.15cm}

We leverage nested virtualization, i.e., one \ac{vm} encapsulates all parts of the testbed environment including the managed host.
This approach allows the entire testbed to be migrated to other hardware platforms for further work and experiments.
We use the KVM hypervisor~\cite{kvm} with QEMU~\cite{qemu} for hardware-assisted virtualization and libvirt~\cite{libvirt} for orchestration.
The managed host, both web servers, and the RADIUS server run as nested \acp{vm} with Ubuntu 17.04~\cite{ubuntu}.
Open vSwitch \cite{openvswitch} serves as SDN switch that is controlled by the Ryu SDN controller \cite{Ryu}.

\subsection{Docker as Container Virtualization Platform for \acp{rac}}
\begin{figure}[htp]
    \centering
    \includegraphics[width=.9\linewidth]{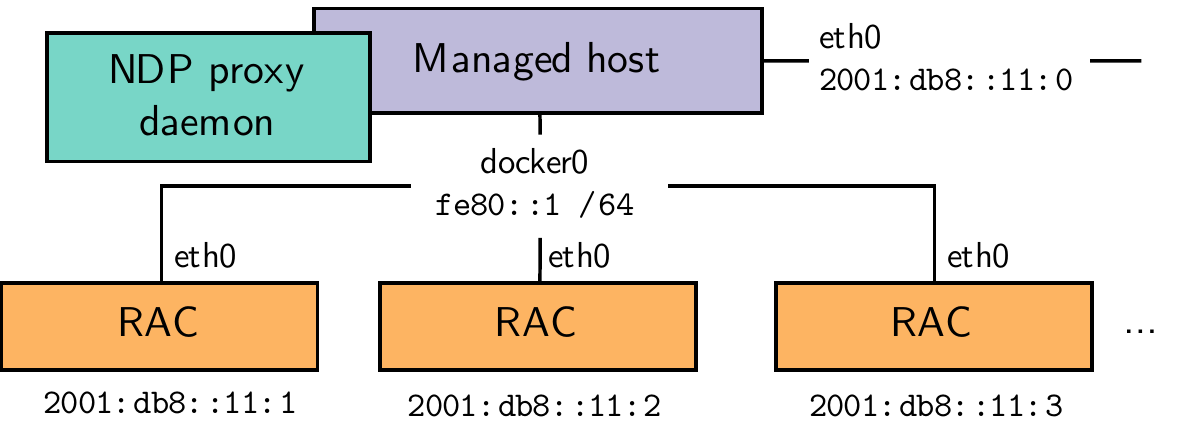}
    \caption{Network configuration of Docker in the testbed environment. Each \ac{rac} gets an IPv6 address of the IPv6 subnet that is assigned to the managed host. The NDP proxy daemon resolves L2 addresses for the \acp{rac}.}
    \label{fig:docker-ipv6}
    \vspace{-0.2cm}
\end{figure}

We use Docker \cite{docker} in Version 17.05 as container virtualization platform to implement \acp{rac}.
We configure the Docker \ac{cmd} so that each \ac{rac} gets a dedicated IPv6 global unicast address that is reachable by other network hosts.
\fig{docker-ipv6} depicts the applied networking configuration that follows the approach presented in \cite{docker-ipv6-cn}.
By default, \acp{rac} only receive a link-local IPv6 address.
Therefore, we set up a fixed IPv6 subnet with routable addresses for \acp{rac}.
The managed host is configured with the IPv6 subnet \texttt{2001:db8::11:0/116} and the \acp{rac} receive an IPv6 address from that range.
The first \ac{rac} receives \texttt{2001:db8::11:1} and the second \ac{rac} \texttt{2001:db8::11:2}, respectively.
The Docker daemon automatically adds routes to the routing table of the system and enables IPv6 forwarding so that all traffic to the IPv6 subnet will be routed via the docker0 interface.
To make the \acp{rac} reachable from other network hosts, we leverage the NDP proxy daemon \cite{ndppd}.
It forwards L2 address resolution for IPv6 addresses of the \acp{rac}, i.e., it listens to neighbour solicitation requests for the \acp{rac} addresses and answers with the MAC address of the managed host. 
Afterwards, packets that address a \ac{rac} are received and forwarded through the Docker host via the docker0 device to the particular \ac{rac}.

\vspace{-0.2cm}

\subsection{802.1X Container Supplicant (802.1X CS)}
\begin{figure}[htp]
    \centering
    \includegraphics[width=.98\linewidth]{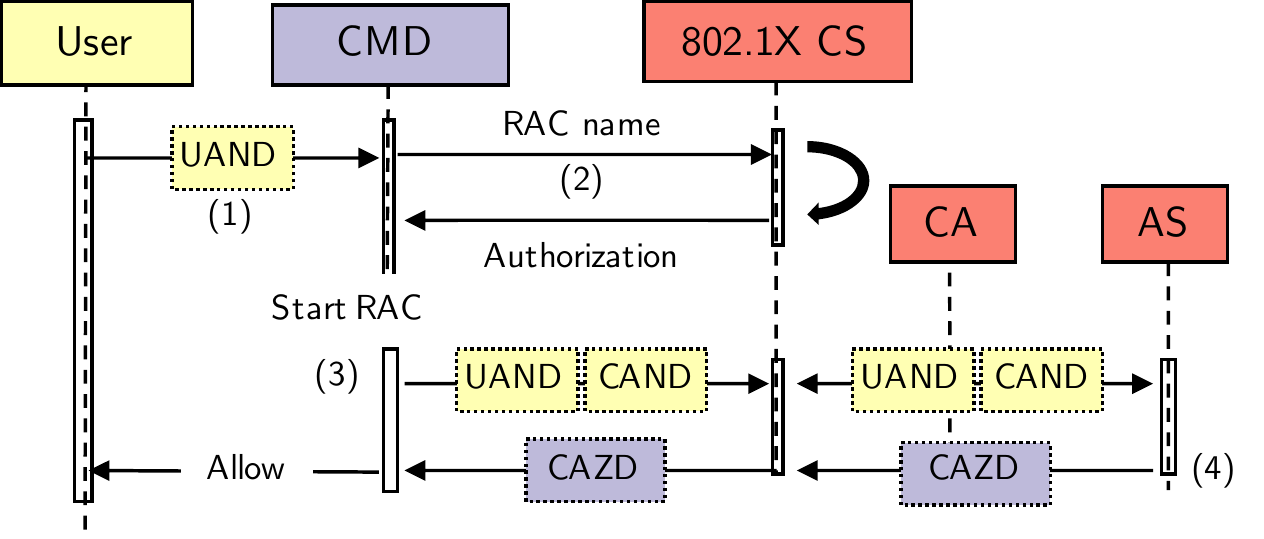}
    \caption{Two-step authorization process in the Docker Authorization Framework \cite{docker-authz}. The \ac{cmd} requests the \ac{8021x-cs} to perform \ac{aa}.}
    \label{fig:authz-process}
    \vspace{-0.3cm}
\end{figure}

We implement the \ac{8021x-cs} as plugin for the Docker Authorization Framework introduced in \sect{container-security-platforms}.
We program the plugin in Python and leverage the Flask \cite{flask} library to implement its REST interface.
\fig{authz-process} depicts the authorization process.
In (1), the user requests the \ac{cmd} to start a container.
The request includes \ac{uand}, e.g., a user name and a password.
The Docker Authorization Framework predefines a two-step authorization process, but we only require the second step.
The first authorization request (2) includes only minimal data, e.g., the name of the \ac{rac} image.
As we solely rely on the second authorization step, the \ac{8021x-cs} responds with a permit by default.
The second authorization request (3) includes \ac{uand} and \ac{cand}.
The \ac{8021x-cs} performs authentication with the \ac{8021x-as} through the \ac{8021x-ca} (3) as discussed before.
In (4), \ac{8021x-as} returns \ac{cazd} that is forwarded to the \ac{8021x-cs} in case of successful \ac{aa}.

\subsection{802.1X Container Authenticator (802.1X CA)}
We implement the \ac{8021x-ca} as SDN application for the Ryu SDN controller framework \cite{Ryu}.
We extend the \ac{8021x-a} of \cite{HaSc17} by adding support for authentication with the \ac{8021x-cs} using \ac{eapoudp}.
The \ac{8021x-ca} opens a UDP socket on port 5995 and waits for connections from the \ac{8021x-cs}.
The \ac{8021x-ca} still may act as legacy \ac{8021x-a} that performs \ac{aa} for network hosts in legacy 802.1X via \ac{eapolan}.
As example for network control with xRAC, we implemented a restricted MAC-learning switch.
It learns MAC addresses from connected hosts but only forwards packets if the IP addresses of both sender and receiver are in a whitelist.
The whitelist contains static entries, e.g., for public servers, and dynamic entries that can be modified by the \ac{8021x-ca} after receiving \ac{cazd} from the \ac{8021x-as}.
We implement the restricted MAC-learning switch by extending the L2 switch \cite{simpleswitch-ryu} from the Ryu SDN controller framework.

\subsection{802.1X Authentication Server}
We leverage the widely-used \ac{8021x-as} software FreeRADIUS and extend its \ac{aa} data model to implement \ac{cand} and \ac{cazd}.
In FreeRADIUS, additional attributes for \ac{aa} and simple policies can be implemented using \acp{vsa} \cite{rfc2865, rfc6158, rfc6929, rfc8044} that are defined with the unlang \cite{unlang} processing language.
The defined \ac{aa} data model can be easily extended and modified by adding more \acp{vsa}.

\subsection{Protected and Public Web Server}
Two web server \acp{vm} in the testbed environment run a Python web server \cite{python-http} that provides HTML files via HTTP.
The protected web server with the static IPv6 address \texttt{2001:db8::aa:0} delivers an HTML page with the sentence "protected content".
The public web server with the static IPv6 address \texttt{2001:db8::bb:0} delivers an HTML page with the sentence "public content".
\section{Experimental Validation}
\label{sec:functional-validation}

We describe the experiment setup and validation experiments for the testbed from \sect{implementation} to validate xRAC.

\subsection{Experiment Setup}
\label{sec:experiment-setup}
The experiments investigate the communication between the managed host, a particular \ac{rac}, the protected web server, and the public web server.
We encapsulate the wget tool \cite{wget} as \ac{rac} to retrieve files using HTTP.
In our experiments, we use the \ac{rac} to request an HTML file from both the protected and the public web server.
We add \ac{uand}, \ac{cand} and \ac{cazd} on the RADIUS server that allows a particular user to run the \ac{rac} and access the protected web server.

\subsection{Validation Experiments}
\begin{figure}[htp]
    \centering
    \includegraphics[width=.94\linewidth]{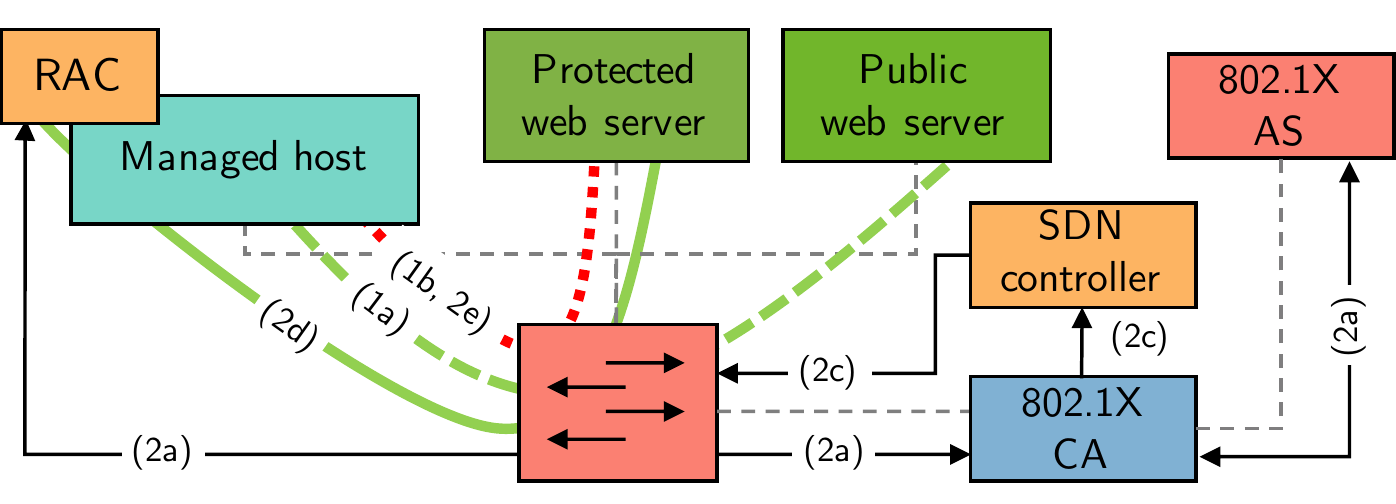}
    \caption{Experiments to investigate the communication between the managed host, a \ac{rac}, and both web servers.}
    \label{fig:experiment}
    \vspace{-0.15cm}
\end{figure}

We perform the following experiments as depicted in \fig{experiment}.
Before launching the \ac{rac}, we validate with ICMP echo requests from the managed host that the public web server (1a) but not the protected web server (1b) is accessible without authorization.
Now, we demonstrate that the integrity of \acp{rac} is verified during authentication, i.e., that a \ac{rac} with a divergent image checksum cannot be started.
We create another \ac{rac} image that encapsulates a patched version of wget and try to start it using the user credentials as set up on the RADIUS server.
Authentication fails, i.e., the \ac{rac} cannot be started on the managed host.
Now, we demonstrate that the correct \ac{rac} can be started and that it can access the protected web server after successful \ac{aa}.
After issuing the command to start the \ac{rac}, it is authenticated and authorized as described before (2a).
The SDN controller receives \ac{cazd} and programs the SDN switch to permit packet forwarding between the \ac{rac} and the protected web server (2c).
Now, the \ac{rac} is able to receive the requested HTML file from the protected web server (2d).
Trying to load the same HTML file with wget directly from the managed host fails (2e), i.e., the protected web server can be reached by the \ac{rac} but not by the managed host.

\vspace{-0.2cm}
\vspace{-0.2cm}

\section{Performance Considerations}
\label{sec:performance-considerations}

Container virtualization has no remarkable performance overhead compared to native application execution \cite{FeFe15}.
xRAC adds delay only to the startup time through the CMD.
The CMD requires time to calculate the container's integrity checksum and to perform AA using network-based 802.1X.
Without xRAC, starting the wget container from \sect{experiment-setup} takes approximately 0.6 s.
With xRAC, the startup time is increased to 1.51 s.
The wget container image is 6 MB large and the calculation of its SHA256 hash takes 0.12 s on our platform.
However, the computation time scales with the image size.
For example, a Chrome browser container image with 880 MB takes 3.2 s.
The duration of the AA operation depends on the performance of the three 802.1X components and the network in between.
The container supplicant is part of the managed host and, therefore, covers only little load.
The container authenticator is part of the controller and responsible for many hosts.
However, its performance can be scaled up by running multiple instances.
The authentication server may be based on RADIUS.
This technology is proven to scale well with large deployments and high load by replicating server instances.
\section{Conclusion}
\label{sec:conclusion}

In this work we proposed xRAC, a concept for execution and access control for restricted application containers (RACs) on managed clients.
It includes authentication and authorization (AA) for RACs such that only up-to-date RAC images can be executed by permitted users.
Moreover, authorization is extended to protected network resources such that authorized RACs can access them.
Traffic control is simplified through the fact that all traffic of a RAC is identified by its IPv6 address.
We presented the architecture of xRAC and showed by a prototype implementation that xRAC can be built from standard technologies, protocols, and infrastructure.
Our prototype of xRAC leverages Docker as container virtualization platform, signalling is based on 802.1X components.
Modifications were needed to the supplicant, the authenticator, and the authentication server so that both user and container AA data can be exchanged.
Moreover, the container authenticator is extended to inform required network control elements about authorized RACs. 
We used the prototype to experimentally validate xRAC and investigate on the performance.
After all, we discussed use cases and showed that xRAC supports software-defined network control and improves network security without modifying core parts of applications, hosts, and infrastructure. 
  
\bibliographystyle{IEEEtran}
\bibliography{paper}
\balance
  
\end{document}